# Complementarity, Augmentation, or Substitutivity? The Impact of Generative AI on the U.S. Federal Workforce

William G. Resh[*,1]
Yi Ming[1,2]
Xinyao Xia[3]
Michael Overton[4]
Gul Nisa Gürbüz[5]
Brandon De Breuhl[6]

Currently Under Review

* Corresponding author: wresh @gsu.edu
[1] Georgia State University, Andrew Young School of Policy Studies
[2] Georgia Institute of Technology, Jimmy and Rosyln Carter School of Public Policy
[3] University of Southern California, Viterbi School of Engineering
[4] University of Idaho
[5] University of California, Santa Barbara
[6] University of Southern California, Sol Price School of Public Policy; RAND Corporation

**Abstract**

How will generative artificial intelligence reshape the competencies required of federal civil servants? This study applies a multi-stage Retrieval-Augmented Generation system, which grounds large language model assessments in curated evidence, to evaluate generative AI's near-term impact on knowledge, skills, and abilities across several hundred federal occupational series. Each occupation is scored along three dimensions: complementarity (AI enhances existing competencies), augmentation (AI requires workers to transform competencies), and substitutivity (AI replaces human tasks). Results reveal that generative AI primarily complements and augments federal roles, with limited substitution potential. Even occupations scoring highest on substitutivity reflect partial automation of routine subtasks rather than position elimination. These findings caution against blunt workforce reductions premised on displacement and instead support targeted reskilling, competency-based planning, and human-AI collaboration. The framework is adaptable to other organizations with structured competency models and job documentation.

**Evidence for Practice**

1. Generative AI's near-term impact on federal occupations is best characterized as task reconfiguration rather than position elimination; complementarity scores consistently exceed substitutivity scores across all occupational families and competency types.
2. Workforce planning strategies should be differentiated by AI impact dimension: tool-adoption training for high-complementarity roles, deeper reskilling for high-augmentation roles, and reallocation toward oversight functions for roles with elevated substitutivity.
3. Even clerical and administrative occupations with the highest substitutivity scores retain essential human functions in exception handling, compliance verification, and accountability for final determinations.
4. Organizations with structured competency models can use this framework to identify where AI is most likely to support work, require reskilling, or automate limited subtasks.
5. Political rhetoric equating AI capability with workforce dispensability conflates complementarity and augmentation with substitutivity, a distinction this analysis clarifies.

## Introduction

The rapid emergence of generative artificial intelligence (AI), including systems capable of producing text, code, and analytical outputs closely resembling human work, has ignited urgent questions about the future of public employment. Political actors have invoked AI as a rationale for sweeping workforce reductions (Tsipursky 2025), while agency leaders grapple with responsible integration. Yet the discourse has outpaced the evidence. Despite a growing literature on AI's labor market effects in the private sector (Eloundou et al. 2024; Frey and Osborne 2017; Brynjolfsson, Li, and Raymond 2025), systematic, occupation-level assessments of how generative AI will affect the competencies defining federal work remain absent.

This gap matters for public administration. The federal executive branch employs over two million civilian workers across an extraordinarily diverse occupational landscape, governed by merit-system principles and standardized competency frameworks (Office of Personnel Management 2018; Kellough and Nigro 2018). Strategic human capital management, long recognized as a cornerstone of effective governance (Ingraham and Getha-Taylor 2004; Selden 2009), depends on accurately understanding which competencies are stable, which must evolve, and which face displacement. Generic assertions that AI will "replace" government workers are analytically impoverished; a granular analysis distinguishing among the different ways AI interacts with human capabilities is required (Roehl and Hansen 2024).

This study addresses this need by developing and deploying a multi-stage retrieval-augmented generation (RAG) analytical system that assesses generative AI's impact on federal occupations at the competency level rather than at the task level that has dominated prior AI-labor analyses (Frey and Osborne 2017; Eloundou et al. 2024). The distinction matters: tasks are the observable activities workers perform, but competencies are the underlying human capacities (i.e., the knowledge, skills, and abilities) from which those tasks are generated (McClelland 1973; Boyatzis 2008).

Because organizations routinely reconfigure task bundles in response to technological and strategic change, task-level findings depreciate rapidly; competencies, by contrast, are more durable properties of occupations and organizational roles, providing a more stable foundation for workforce planning (Spencer and Spencer 1993; Campion et al. 2011). The system extracts predominant competencies from authoritative job documentation, then evaluates each competency along three theoretically distinct dimensions: *complementarity* (AI enhances existing human competencies without altering the worker's role), *augmentation* (AI requires workers to transform how they exercise competencies), and *substitutivity* (AI replaces human task performance). This competency-level taxonomy moves beyond the binary automation framing and offers a more durable basis for workforce planning. The three-part taxonomy also structures the model prompts used in the analysis, ensuring that assessments distinguish among support, transformation, and substitution rather than collapsing them into a single automation judgment

The central finding is that generative AI's near-term impact on the federal workforce is best characterized as *reconfiguration rather than replacement.* Across all occupational families, complementarity scores are highest (mean = 3.56), augmentation scores are moderate (mean = 3.31), and substitutivity scores are consistently lower (mean = 2.38). Even in occupations where substitutivity is comparatively elevated (principally clerical and administrative roles), assessments indicate partial automation of routine subtasks rather than position elimination. Occupations demanding real-time judgment, interpersonal engagement, and physical accountability show minimal substitution potential. These findings challenge the political rhetoric of AI-driven mass displacement and instead support workforce planning centered on human–AI collaboration.

The federal government serves as a prototype because its size, occupational diversity, transparency, and standardized competency frameworks make it ideal for developing and testing this methodology (Rainey and Bozeman 2000; Attard et al. 2015). The methodology is replicable: any organization with established competency models can apply a similar approach. The article proceeds by developing the theoretical rationale for the three-dimensional impact framework. We then describe the methodology and present distributional results, occupational rankings, and a detailed illustrative case. Finally, we discuss the implications for strategic human capital planning, agency-level workforce development, and the broader governance of AI integration in government.

We present this RAG-based framework as a prototype methodology. This is a first-generation model whose primary contribution is demonstrating the feasibility and analytical value of applying retrieval-augmented LLMs to occupational impact assessment, rather than offering definitive forecasts. While the results presented here are promising and substantively consistent with emerging findings in labor economics, we acknowledge that comprehensive validation that includes multi-model replication, formal reliability and validity testing, and entailment evaluation (Overton, Robison, and Sheneman 2025) remains essential and is the subject of ongoing and future work.

## Theoretical Framework

### AI and the Transformation of Public Work

The scholarship on AI and employment has evolved through several phases. Early analyses focused on binary automation risk, estimating the probability that entire occupations would be replaced (Frey and Osborne 2017). More recent work recognizes that technology rarely eliminates whole jobs but instead reshapes task composition, automating some activities while creating or enhancing others (Autor 2015; Acemoglu and Restrepo 2019). Generative AI intensifies this dynamic because its capabilities extend into knowledge work, including writing, analysis, coding, and content creation, which was previously considered resistant to automation (Eloundou et al. 2024; Brown et al. 2020). Bommasani et al. (2021) coined the term "foundation models" to describe large-scale AI models like GPT-3 that serve as a base for diverse applications, highlighting both opportunities and risks for job displacement, ethical considerations, and the need for human oversight.

For public administration, these developments intersect with longstanding concerns about workforce capacity and governance. The federal workforce embodies institutional competencies through which the government exercises discretion, ensures accountability, and delivers public value (Perry and Hondeghem 2008). Strategic human capital planning requires understanding whether a task *can* be automated and, separately, whether it *should* be, given the requirements of democratic governance, public trust, and regulatory accountability (Young, Bullock, and Lecy 2019; Busuioc 2021; Wirtz, Weyerer, and Geyer 2022). AI and automation may lead to large-scale reconfiguration of job tasks, requiring workers to adapt their skills and competencies (Autor et al. 2022). While some tasks might be automated, new tasks and roles will emerge that necessitate continuous learning and adaptability. Nedelkoska and Quintini (2018) analyzed OECD countries and found that while automation poses risks, it also offers opportunities for job enhancement and creation.

Recent policy developments underscore this urgency. Under President Biden, the Office of Personnel Management emphasized ethical AI use aligned with existing competency frameworks, encouraging agencies to ensure that automation augments human contributions. Federal agencies were directed to align AI applications with existing competency frameworks through OPM's guidance on the use of generative AI for the federal workforce, including use-case inventories and newly developed AI training programs. Under President Trump, Executive Order 14179 dismantled prior mandates on "safe and trustworthy" AI, championing instead a doctrine of accelerated adoption. The Office of Management and Budget's (2025) new directives empower agencies to pursue AI integration with unprecedented autonomy, provided they uphold basic privacy and civil liberty standards. Whatever the governing rationale, generative AI is now moving into core administrative settings, making competency-level analysis of public work increasingly important.

## Competency Modeling and Natural Language Processing

Competency modeling is the established approach to aligning employee capabilities with organizational goals amid technological change (Boyatzis 2008; Shippmann et al. 2000). Competencies, as defined by OPM, are measurable patterns of knowledge, skills, and abilities (KSAs) essential for successful job performance (Office of Personnel Management 2018). The choice to analyze competencies rather than tasks is deliberate and analytically consequential. The dominant tradition in AI-labor scholarship takes the task as the unit of analysis, decomposing occupations into discrete activities and estimating each activity's susceptibility to automation (Frey and Osborne 2017; Autor, Levy, and Murnane 2003; Felten, Raj, and Seamans 2021; Eloundou et al. 2024). Task-level analyses have yielded important insights, but they suffer from a fundamental limitation: tasks are transient composites that organizations routinely reconfigure in response to technological, regulatory, and strategic changes (Autor 2015). What persists across such reconfigurations are the underlying competencies from which tasks are generated. As McClelland (1973) argued in his foundational critique of aptitude testing, it is competencies that predict sustained performance across varying work contexts, not the specific activities through which they are expressed. Subsequent scholarship has reinforced this insight: competencies function as generative capacities that enable workers to produce, adapt, and recombine tasks as role demands shift (Boyatzis 2008; Spencer and Spencer 1993; Shippmann et al. 2000). In the language of Autor, Levy, and Murnane's (2003) task-based production framework, competencies are the latent inputs from which the "task content" of an occupation is composed and recomposed. Analyzing AI's impact at the competency level therefore captures a comparatively more durable structure of human capital within occupations and organizational roles, providing workforce planners with a more stable foundation for investment that does not depreciate as rapidly with each technological iteration (Rodriguez et al. 2002; Campion et al. 2011). This approach is consistent with competency-based human resource management, which emphasizes aligning individual capabilities with organizational goals precisely because competencies are comparatively more stable and transferable than the specific task bundles they underwrite (Ulrich and Dulebohn 2015). Traditional competency models, however, may not adequately account for the dynamic nature of AI advancements. As Sparrow and Makram (2015) note, organizations must develop dynamic competencies that enable adaptability and innovation. The very development and training of skills in the labor market is endogenous to the rapid evolution in sophistication and capacity of generative AI itself; humans must adapt to the technology's increased capacity as generative AI adapts to human learning and capacity to use the technology in turn (Kim 2024).

Generative AI accentuates the need for new competencies and evolution of existing ones. Augmentation strategies necessitate competencies in AI literacy and collaboration with intelligent machines (Davenport and Kirby 2016). Wilson and Daugherty (2018) identify "fusion skills" required for human-AI partnerships, including responsible normalization and reciprocal apprenticing. The World Economic Forum (2025) identifies analytical thinking, active learning, and technology use as key future competencies. These competencies are increasingly important as generative AI automates routine tasks, shifting the focus to higher-order skills in problem-solving and ethical reasoning (Hecklau et al. 2016; Felzmann et al. 2020).

Advancements in natural language processing (NLP), particularly transformer-based models, have made it possible to extract, represent, and compare competencies from unstructured text with unprecedented precision. The introduction of the Transformer architecture by Vaswani et al. (2017)

paved the way for models like BERT (Devlin et al. 2019) and GPT-3 (Brown et al. 2020), which offer contextualized word embeddings capturing nuanced meanings based on context. Applying these techniques to analyze federal job descriptions enables the capture of subtle differences in competencies across occupations and the identification of which tasks are susceptible to various forms of AI impact.

**Beyond Automation: A Three-Dimensional Taxonomy**

A single automation-risk score is insufficient for workforce planning because it conflates fundamentally different modes of human–AI interaction. Drawing on established distinctions in the technology-and-work literature (Autor 2015; Brynjolfsson and McAfee 2014), this study proposes a taxonomy of three dimensions. Critically, this taxonomy is applied at the competency level rather than the task level. Prior frameworks that decompose occupations into tasks and then score each task's automation probability (Frey and Osborne 2017; Eloundou et al. 2024) answer an important but narrow question: which current activities can machines perform? A competency-level taxonomy asks the antecedent question: how does AI interact with the human capacities that generate, adapt, and govern those activities? Because competencies are the generative inputs from which task performance flows (McClelland 1973; Boyatzis 2008), assessing AI's impact at this level yields insights that remain valid even as the specific tasks composing an occupation are reorganized.

***Complementarity.*** Complementarity is the ability of generative AI to work alongside humans by enhancing capabilities through distinct AI strengths without replacing human labor (Autor 2015; World Economic Forum 2025). These systems handle data-heavy or repetitive tasks, freeing humans for higher-order thinking, judgment, and decision-making. For example, the Census Bureau's BEACON (Business Establishment Automated Classification of NAICS) system uses machine learning to help Economic Census respondents classify their business establishments into six-digit NAICS industry codes in real time. BEACON takes respondent-provided text descriptions as input, applies natural language processing and information retrieval techniques trained on past Economic Census responses and other authoritative data sources, and returns a ranked list of candidate industry codes for the respondent to select from. During the 2022 Economic Census, the tool was used over half a million times, significantly reducing the clerical burden of manual industry code assignment and allowing analysts to concentrate on cases requiring substantive economic judgment (Dumbacher et al. 2025).

***Augmentation.*** Augmentation involves integrating AI with human intelligence to create synergy, enabling better decisions and improved productivity. Unlike complementarity, augmentation requires humans to change their orientation to a given competency by understanding how to integrate AI into that competency (Brynjolfsson and McAfee 2014; Brynjolfsson, Li, and Raymond 2025). The Department of Energy's AI Strategy identifies multiple AI tools deployed across its operational infrastructure, including the Data Analytics and Machine Learning (DAMaL) tools, which support analysis, data management, and process automation across DOE sites, field offices, and headquarters, as well as separate Project Management AI tools, such as the Project Assessment and Reporting System (PARS) AI Assistant, which provides real-time insights, advanced document indexing, and AI-assisted project assessment and compliance review (Department of Energy 2025). These systems illustrate augmentation in practice: staff must learn to integrate AI-generated analytics into their existing workflows and professional judgment, adjusting how they exercise competencies in project oversight and technical analysis while retaining authority over final decisions.

***Substitutivity.*** Substitutivity refers to AI systems designed to fully automate tasks humans traditionally perform, particularly in areas where human cognition is less critical, such as routine, repetitive tasks (Frey and Osborne 2017; Eloundou et al. 2024). Customs and Border Protection's Integrated Defense and Security Solutions (IDSS) system illustrates how AI can automate detection functions previously performed by human operators. The system combines sensor data from radio frequency detections, radar, and infrared/electro-optical cameras, using AI to fuse these feeds and provide detections on a user interface for further agent investigation. Critically, the system does not autonomously mitigate threats; it detects and provides a digital track for human review, shifting operator effort from continuous monitoring of surveillance feeds to evaluating AI-flagged alerts (Department of Homeland Security 2025).

These dimensions are conceptually distinct yet interact in practice within a single occupation or competency. AI may simultaneously complement some aspects of a knowledge base, require augmented skills, and substitute for routine subtasks within the same job series. Each competency is therefore evaluated along all three dimensions independently. The balance among these three dimensions will shape future labor markets, with each bringing distinct implications for job complexity, human involvement, and task automation (Frey and Osborne 2017; Brynjolfsson and McAfee 2014; Eloundou et al. 2024). Table 1 summarizes the taxonomy.

*[Table 1 here]*

**Why the Federal Government?**

The U.S. federal government serves as an ideal prototype for several reasons. Its scale, with approximately two million civilian employees across hundreds of occupational series distributed across geographic and economic regions, provides the breadth needed to examine AI's differential effects. OPM's standardized competency frameworks, the Federal Workforce Competency Initiative, and merit system practices provide structured descriptions essential for systematic analysis (Office of Personnel Management 2018). The government's comparative transparency enhances feasibility (Attard et al. 2015), and innovations in public sector workforce practices frequently inform private sector approaches (Rainey and Bozeman 2000), where the proprietary and internal nature of their workforce strategies and assessment analytics may be less transparent. The methodology is replicable across other institutions by training domain-specific models on organization-specific job descriptions and competency frameworks (Campion et al. 2011; Ulrich and Dulebohn 2015).

## Methodology

### Overview

The methodology proceeds in two stages, both powered by a multi-stage Retrieval-Augmented Generation (RAG) system. RAG is an AI architecture that combines the strengths of information retrieval and generative models to provide more accurate and context-rich responses (Lewis et al. 2020). Pre-trained language models can learn a substantial amount of knowledge from training data, but their knowledge is limited to a static training database and is not updatable. Such inflexibility often leads to "information hallucination," which occurs when the model generates outputs that are not grounded in its training data (Hadi et al. 2023). RAG addresses these limitations by first using a retrieval model to search a large collection of documents and extract relevant information as input, which is then fed into a generative model to create a coherent and informative output. Such input acts as a non-parametric memory that is updatable (Izacard et al. 2022). For public administration researchers, the key insight is that RAG functions much like a research assistant who consults a library of relevant documents before answering a question, rather than relying on recall alone.

In Stage 1, the system processes official job descriptions, classification standards, and competency frameworks to identify the three predominant knowledge areas, three skills, and three abilities for each occupational series, yielding nine competency units per occupation. In Stage 2, a separate RAG system evaluates each KSA along the three dimensions of complementarity, augmentation, and substitutivity, producing a 1 to 5 Likert score and a narrative justification for each assessment. Each dimension is evaluated through an independent model call to prevent anchoring bias, the tendency for an initial assessment to bias subsequent ones (Stureborg, Alikaniotis, and Suhara 2024). The system is implemented using LangChain, an open-source framework designed to simplify the implementation of RAG in practical applications (LangChain 2024). Figure 1 diagrams the complete methodology.

*[Figure 1 here]*

**Knowledge Bases**

The Stage 1 knowledge base supports competency extraction and integrates the O*NET KSA taxonomy as the foundational competency framework, incorporating occupational KSA databases, crosswalk files, federal job descriptions aligned with OPM's Handbook of Occupational Groups and Families (Office of Personnel Management 2018), Federal Position Classification Standards (Office of Personnel Management 2009), and validated competency ratings from the Federal Workforce Competency Initiative (Office of Personnel Management 2023). O*NET datasets (National Center for O*NET Development 2025) provide more granular information on occupational competencies, their importance, and required proficiency levels. By incorporating these structured sources, the system can better contextualize the unique demands and qualifications associated with each position.

The Stage 2 knowledge base provides the empirical and theoretical foundation for AI impact assessment. Following Davenport and Paredes (2025) as a structured map of the forecasting landscape, the corpus draws from occupation-level automation risk estimates (Frey and Osborne 2017), task-based exposure frameworks (Felten, Raj, and Seamans 2018, 2021), generative AI capability assessments (Eloundou et al. 2024), task-based production theory (Autor, Levy, and Murnane 2003; Acemoglu and Restrepo 2019), and policy sources (Congressional Budget Office 2023; World Economic Forum 2025). The knowledge base also integrates ethical and regulatory frameworks relevant to AI adoption in the public sector, including literature on sector-specific constraints and ethical principles (Felzmann et al. 2020). This diversity ensures retrieval does not privilege a single forecasting paradigm. The knowledge base defines the analytical boundary within which the model reasons: assessments are conditioned on the paradigms embedded in the retrieval corpus.

**Document Processing and Vector Database**

Documents are divided into 500-character segments with 100-character overlap using a two-pass chunking strategy. The first pass splits by newline boundaries respecting the 500-character target, and a second pass further subdivides any chunks exceeding the limit through fixed-length segmentation. This two-pass approach ensures that no chunk exceeds the target length, preserving precision of knowledge while maintaining semantic coherence (Zhong et al. 2024). Segments are converted into dense vector representations using OpenAI's embedding API and stored in a FAISS (Facebook AI Similarity Search) database optimized for similarity search and clustering (Johnson, Douze, and Jégou 2019). By training on a corpus of federal job descriptions and competency frameworks, domain-specific embeddings are created that reflect the unique language and nuances of federal occupations, addressing limitations of generic embeddings that may overlook important contextual details (Levy and Goldberg 2014). The FAISS vector store is constructed directly from the text chunks and their corresponding embeddings, enabling rapid nearest-neighbor lookups that preserve nuanced semantic relationships. These chunking and embedding decisions are particularly important for occupational text, where competency descriptions contain dense, domain-specific terminology that generic segmentation could fragment across chunks and sever the link between a job and its operational context.

**Model Specification**

The primary LLM was OpenAI GPT-5.2, accessed February 20, 2026. Each job in both stages was treated as an individual input into the RAG model. The model's temperature was set at the default of 1.0. In Stage 1 (KSA extraction), prompt token usage varies substantially due to differences in job description length. Empirical logs show that Stage 1 prompt tokens average approximately 11,400 tokens per request, with values ranging from roughly 2,300 to 45,500 tokens. In Stage 2 (AI impact scoring), a structured prompt is applied to the extracted KSAs without large contextual inputs, resulting in more stable token usage, typically between 2,180 and 2,390 prompt tokens per request. To assess cross-model robustness, we ran the full pipeline on both GPT-4o and GPT-5.2 and found similar scoring patterns and rank ordering (GPT-4o accessed on February 20, 2026, with a 128,000-token context window and GPT-5.2 with a 400,000-token context window).

### Prompt Design

Each CAS dimension prompt is structured with the following components, in order: (1) a role assignment establishing the model as an expert analyst specializing in AI impact on federal government occupations; (2) a context frame specifying that impact is assessed across multiple underlying dimensions and identifying the specific dimension under evaluation; (3) a precise definition of the dimension; (4) a task specification listing the nine extracted KSAs for the occupation; (5) instructions specifying the 1 to 5 Likert scale with labeled anchors and requesting a brief justification for each score; and (6) a strict JSON output schema. A zero-shot prompt strategy with separate calls for each dimension prevents anchoring bias (Stureborg, Alikaniotis, and Suhara 2024). Neutral language avoids priming toward optimistic or pessimistic assessments (Atreja et al. 2024; Malberg et al. 2024). Explicitly instructing models to analyze the provided context and provide an explanation for their decisions increases compliance with instructions and results in outputs more closely correlated with human outputs (Chiang and Lee 2023). Complete prompts appear in appendix A.

## Results

### Cross-Model Consistency Check

The RAG pipeline was run on both GPT-4o and GPT-5.2 to assess whether AI impact scores reflect genuine structural features of occupations rather than artifacts of any single model. Figure 2 presents ridgeline plots of score distributions across KSAs for both models. For each job series, an average score was created using all nine individual KSA scores within each AI impact category. The analysis covers 435 occupational series (308 white-collar, 127 trade, craft, and labor).

*[Figure 2 here]*

Both models converge on the same fundamental finding: complementarity highest, augmentation moderate, substitutivity lowest. Rank ordering of occupational series is largely preserved and distributional shapes are similar, suggesting that the scoring framework appears to capture stable relationships between generative AI and federal occupations. Where the two models diverge, GPT-5.2 demonstrates heightened sophistication in narrative reasoning. For example, when assessing the complementarity of Analytical Skill in the 0301 (Miscellaneous Administration and Program) series, GPT-4o offers a general statement about AI aiding data manipulation; GPT-5.2 specifies that AI can synthesize program information and propose analytical frameworks while clarifying that humans remain central for source verification and high-stakes judgment. For augmentation of Management Principles in the 0343 (Management and Program Analysis) series, GPT-4o broadly notes that AI optimizes processes; GPT-5.2 identifies concrete subtasks, including interview guides, extraction checklists, and document review protocols, and describes how the skill shifts toward AI-assisted triage plus rigorous verification. For substitutivity of Negotiation Techniques in the 1102 (Contracting) series, GPT-5.2 assigns a lower score (2 vs. 3), reasoning that while AI can prepare strategies and generate alternative language, real-time bargaining, credibility signaling, and relationship management remain fundamentally human. Given GPT-5.2's stronger reasoning performance, its outputs are used for the remainder of the analysis. For the reader's own assessment of the construct validity of our system's analyses and reasoning narratives (using both 4.0 and 5.2), our data are available upon publication.

**AI Impact Distributions**

Descriptive analysis evaluates AI impact scores across three dimensions for Knowledge, Skills, and Abilities, distinguishing white-collar and trade, craft, and labor (TCL) occupational categories. White-collar roles often involve information-based, analytical, and data-intensive tasks, making them more likely to benefit from generative AI's complementarity and augmentation capabilities. TCL occupations are more manual-task-oriented, where generative AI's substitutive potential may be relatively higher but is still limited by the nature of physical and contextual work. This distinction ensures that workforce strategies and policy recommendations are tailored to the specific needs of each occupational group. Figures 3 and 4 present distributional plots; tables 2 and 3 provide summary statistics.

*[Figure 3 here]*

*[Figure 4 here]*

*[Table 2 here]*

*[Table 3 here]*

This pattern is particularly evident when examining occupations that combine relatively high complementarity and augmentation with low substitutivity. These roles illustrate the modal case in the data: AI meaningfully enhances task efficiency and reshapes workflows, while preserving the need for human labor. Analytical and program-oriented occupations benefit from AI's ability to synthesize information and structure outputs, yet still rely on human judgment for validation and strategic application. Similarly, even in more procedural or routine roles, AI tends to automate discrete subtasks while preserving human responsibility for oversight and final decision-making. High complementarity and augmentation scores in roles like engineering, conservation, and skilled trades show AI's potential to enhance efficiency and decision-making. The persistently low substitutivity scores in these professions reinforce the irreplaceable value of human dexterity, adaptability, and field-based expertise.

Complementarity, where AI enhances human capabilities without replacing them, had the highest mean scores across all KSAs (knowledge: 3.45; skills: 3.39; abilities: 3.31). Augmentation was moderate (2.91 to 3.10). Substitutivity was lowest (2.40 to 2.87). White-collar occupations exhibited higher complementarity (mean 3.46) and augmentation (mean 3.13) than TCL occupations (3.18 and 2.77, respectively). Substitutivity was consistently low across both categories (2.66 and 2.50), underscoring generative AI's supportive role. The high complementarity scores confirm the projection that generative AI will predominantly function as a collaborative tool, particularly in roles requiring data analysis and decision support. The moderate augmentation scores support the expectation that human workers will need to adapt their KSAs to leverage AI effectively, while the low substitutivity scores reinforce the idea that full automation remains unlikely for most federal occupations.

**Ranking Occupations by AI Impact**

Figures 5 through 7 rank white-collar occupational series by average scores for each dimension, revealing which jobs are influenced most and least by generative AI. This comparative analysis underscores that AI's impact varies significantly across white-collar occupations, with certain roles poised for high or low levels of each dimension.

***[Figure 5 here]***

***Substitutivity.*** The highest substitutivity scores appear in clerical roles such as voucher examining, secretary, and miscellaneous clerk and assistant, where AI can draft routine correspondence, extract and classify documents, compare records at scale, and flag exceptions for human review. These roles involve procedural, high-volume tasks that AI can largely handle independently, with human involvement shifting toward exception handling, compliance verification, and accountability for final determinations. In contrast, aircraft operation, wildlife refuge management, and recreation aid exhibit the lowest substitutivity. These roles require real-time physical control, on-site coordination, and accountable human judgment in dynamic environments. AI may offer decision-support prompts, documentation assistance, or procedural guidance, but the embodied, interpersonal, and accountability-driven nature of these roles makes substitution severely limited.

*[Figure 6 here]*

***Augmentation.*** Writing and editing, economist, and voucher examining series all benefit significantly from AI's ability to accelerate drafting, synthesize literature, and cross-reference records. The human role shifts accordingly: writers evolve toward directing AI outputs and ensuring narrative accuracy, economists toward model selection and causal interpretation, and examiners toward validating AI-flagged discrepancies and audit defensibility. Voucher examining ranks high on both augmentation and substitutivity, reflecting a role where AI simultaneously automates routine tasks and enhances the examiner's remaining analytical responsibilities. Medical technician, customs and border protection interdiction, and aircrew technician show limited augmentation. While AI can support these occupations through procedural checklists, training aids, and post-task analysis, it does little to transform the core competencies: performing lab tests remains protocol-bound and hands-on, interdiction hinges on real-time threat assessment, and aircrew diagnostics depend on situational awareness under time pressure.

*[Figure 7 here]*

***Complementarity.*** Investigative analysis, health aid and technician, and data science exemplify high complementarity, with AI accelerating research, surfacing relevant references, and proposing analytical approaches while human workers retain responsibility for judgment, verification, and accountability. Unlike augmentation, which transforms how specific tasks are performed, complementarity captures AI's role as a supportive backbone for human-led processes: investigators determine which leads merit pursuit, technicians perform procedures and ensure patient safety, and data scientists validate statistical rigor and causal reasoning. Police and security guard, medical technician, and fire protection exhibit limited complementarity, as these roles demand real-time judgment, interpersonal engagement, and physical accountability in high-stakes environments.

**Occupations with Minimal AI Impact**

Several occupations exhibit minimal scores across all three dimensions, highlighting roles where AI offers limited contribution in any form. Nuclear Materials Courier (Series 0084) receives complementarity and augmentation scores of 2 and a substitutivity score of 1, reflecting irreducible demands of physical security, real-time threat assessment, and strict regulatory compliance. AI can assist in route optimization and security monitoring, but the task itself remains deeply human-driven. Cash Processing (0530) similarly scores 2, 2, and 1, as handling financial transactions, verifying signatures, and reconciling discrepancies require high scrutiny and regulatory compliance. Language Clerical roles (1046) score low in substitutivity (2) due to the complexity of linguistic nuances, cultural understanding, and context-driven interpretation required in translating and clerical linguistic work. Position Classification (0326) also ranks among the lowest, as assessing organizational structures, policy compliance, and personnel decisions requires human judgment AI cannot currently replicate.

The findings across occupational rankings reinforce a clear conclusion: while AI serves as a powerful tool in data-driven and knowledge-intensive fields, its impact varies significantly across professions. As AI continues to evolve, agencies must adopt a balanced integration strategy that leverages AI for its strengths in data analysis and automation while ensuring that essential human skills remain central in physically demanding, socially interactive, and judgment-based roles. Future workforce planning should emphasize AI-augmented skill development without assuming full automation in areas where human presence remains indispensable.

**Applying the Framework: The Case of Federal Economists**

The Economist series (0110) illustrates the framework's granularity. Figure 8 presents the heatmap of AI impact scores across all nine KSAs.

*[Figure 8 here]*

Statistical Theory and Data Analysis receive the highest complementarity scores (5), reflecting AI's capacity to accelerate data manipulation and statistical computation. The system's narrative justification states that AI enables faster and more precise handling of large datasets, complementing human skills in interpretation and decision-making. Data Analysis receives a substitutivity score of 4, as AI excels at automated statistical analysis, yet Economic Principles receives complementarity of 4 but substitutivity of only 2, because interpreting economic principles requires contextual knowledge and adaptability that AI currently lacks.

Economic Institutions received a complementarity of 3 and augmentation of 2. The moderate complementarity was justified in terms of human judgment: AI can assist in understanding economic institutions by providing comprehensive data analysis, though human judgment is crucial for context and insight. The lower augmentation score was justified according to a separate aspect of that knowledge base, emphasizing the importance of contextual understanding: while AI can assist in gathering and organizing information about economic institutions, the nuanced understanding and contextual analysis still heavily rely on human knowledge.

These nuances illustrate the multifaceted nature of AI's impact on occupational competencies. AI's contributions operate along a spectrum where complementarity, augmentation, and substitutivity interact in complex ways. Even within a single competency, AI's ability to process vast amounts of data may serve as a strong complement to human decision-making, while its limited capacity for contextual reasoning prevents it from fully substituting human expertise. A single automation index would obscure these distinct strategic implications. Appendix D demonstrates the consistency of these findings with emerging scholarship on AI's impact in economics.

**Evaluation Approach**

This research intersects two LLM task categories: LLM-as-a-judge, where LLMs evaluate natural language tasks without reference answers (Zheng et al. 2023), and Natural Language Inference (NLI), emphasizing logical reasoning rather than factual accuracy (Yu et al. 2024). LLM assessments are scalable, explainable, and align with human assessments (Hu et al. 2024; Desmond et al. 2024). This study extends the approach by requesting explanations behind LLM decisions, which improves consistency (Atreja et al. 2024) and provides auditable data.

The NLI-generated justifications provide construct plausiblity, revealing how complementarity, augmentation, and substitutivity overlap in real-world application. The key goal is determining entailment (Bowman et al. 2015): does the explanation logically follow from the assigned score? This provides assessment that traditional accuracy metrics cannot fully capture, as human opinions may not represent a useful standard for forward-looking projections. Quality evaluation criteria focus on assessing readability (quality of the output's language elements) and adequacy (quality of the output's task elements) of LLM outputs (Hu et al. 2024). These quality evaluations provide additional robustness checks that can also be compared across outputs from other LLMs. Future work will systematically evaluate output reliability and validity rooted in measurement theory (Wallach et al. 2024; Overton, Robison, and Sheneman 2025), including entailment testing and quality evaluation, and will expand the analysis across multiple open-source LLMs for inter-rater reliability.

## Discussion

### Reconfiguration, Rather than Replacement

The findings converge on a central conclusion: generative AI's near-term impact on the federal workforce is best understood as competency reconfiguration rather than position elimination. AI shifts how competencies are exercised within occupations, such as amplifying productivity in some domains, introducing new modes of human–AI collaboration in others. But its substitutive impact on the underlying human capacities through which effective public service operates is far more limited. Because this study assesses impact at the competency level rather than the task level, the findings capture patterns that task-based analyses are poorly positioned to detect. Task inventories change as organizations adopt new tools and workflows; the competencies evaluated here (e.g., analytical reasoning, regulatory knowledge, interpersonal judgment) are more durable properties of occupational roles (Boyatzis 2008; Shippmann et al. 2000). The persistence of high complementarity and moderate augmentation scores across competency types suggests that AI is reshaping how federal workers deploy their capabilities, not rendering those capabilities obsolete. Indeed, algorithmic systems in government require many of the very competencies our framework identifies as durable (Gules-Guctas 2026). This empirical pattern provides substantial evidence against generalized narratives of widespread automation and supports a more nuanced view centered on human–AI complementarity and adaptive competency evolution.

This finding is particularly consequential given proposals for wholesale workforce reductions justified by assertions about AI's displacement potential (Tsipursky 2025). The analysis suggests such proposals rest on an empirically unfounded conflation of AI's capacity to complement and augment human work with its far more limited capacity to substitute for it. Policymakers who treat AI as a straightforward replacement for human labor risk degrading institutional competencies essential for effective governance.

**Implications for Strategic Human Capital Management**

The three-dimensional framework yields differentiated strategic recommendations. For high-complementarity occupations, the priority is ensuring workers have access to AI tools and the infrastructure to use them; training should focus on tool adoption. For high-augmentation occupations, agencies should invest in reskilling programs helping workers integrate AI-generated outputs into professional judgment, including training in AI literacy, output validation, and collaborative workflows (Davenport and Kirby 2016; Wilson and Daugherty 2018). For occupations with elevated substitutivity, workforce planning should anticipate reallocation from routine subtasks to oversight, exception handling, and quality assurance.

These recommendations are occupation-specific and competency-specific. A one-size-fits-all approach is inconsistent with the evidence. The dynamic and iterative model addresses limitations identified in prior research by incorporating detailed competency data and accounting for the evolving capabilities of generative AI. By continuously updating according to AI developments and changing competency requirements, the model enables organizations to develop targeted strategies for reskilling and upskilling employees in a timely manner. The findings also emphasize the importance of developing complementary competencies that AI cannot replicate, such as critical thinking, ethical judgment, and complex problem-solving. By acknowledging the continuous evolution of AI and its impact on required competencies, the model aligns with the concept of human-AI collaboration, where AI augments human capabilities.

**The Federal Government as a Workforce Laboratory**

The federal government as prototype demonstrates both the feasibility and value of competency-level AI impact assessment at institutional scale. Any organization with established competency models and structured job descriptions can apply similar techniques, substituting their own frameworks and knowledge bases. By training domain-specific embeddings on industry-relevant job descriptions, organizations capture unique linguistic nuances (Levy and Goldberg 2014). The customization of AI agents can be further enhanced by incorporating industry-specific technological trends and economic factors, ensuring that the methodology aligns with regulatory and ethical considerations pertinent to each organization's context. The federal government's unique combination of scale, transparency, and standardization positions it as a natural laboratory for testing workforce analytics that can then diffuse to other sectors (Rainey and Bozeman 2000). The multi-level analysis allows for agency-level insights, enabling tailored recommendations that address the unique challenges and priorities within different federal organizations.

**Limitations and Future Research**

Several limitations warrant acknowledgment. First, assessments are synthetic: they represent LLM judgments grounded in curated evidence rather than direct measurements of AI's actual impact on federal workers' daily activities. While narrative justifications improve interpretive transparency and cross-model comparison provides a useful robustness check, scores should be understood as informed projections. Future work should integrate survey and interview data from federal employees, experimental studies of AI adoption, and longitudinal competency tracking.

Second, assessments are conditioned on the retrieval corpus. Because the corpus privileges task-based production theory, the system may be more sensitive to task-level reallocation than occupation-level displacement. Future iterations could incorporate alternative paradigms or adversarial knowledge bases.

Third, the analysis operates at the occupational series level and does not capture within-series variation across grade levels, agencies, or geographic contexts. Extending to position-level data, running across multiple open-source LLMs for inter-rater reliability, and systematic evaluation of output validity rooted in measurement theory (Wallach et al. 2024; Overton, Robison, and Sheneman 2025) are natural next steps. A multi-model approach will shed light on how variations in architectural design, training data composition, and ethical alignment protocols influence model behavior, helping distinguish between universal trends in LLM reasoning and idiosyncrasies tied to individual models. The findings presented here establish a critical baseline, offering a reference point for tracking the evolution of LLM capabilities as the technology advances.

## Conclusion

The study also highlights the importance of refining semantic representations through validated competency and task ratings, which serve as a benchmark for significance and proficiency levels. By comparing findings with competency models from other sectors, the broader applicability of the approach is suggested. The multi-level analysis allows for agency-level insights, enabling tailored recommendations that address the unique challenges and priorities within different federal organizations. This level of customization underscores the potential for the methodology to be applied broadly, offering a valuable tool for organizations seeking to navigate the evolving landscape of AI integration and workforce transformation. Ultimately, this research provides insights that can be applied across various sectors, enhancing the adaptability and effectiveness of AI integration strategies. By embracing this approach, organizations can better align their workforce development with the dynamic needs of their respective industries, fostering a more resilient and adaptable workforce in the face of technological change.

This study presents a competency-based framework that informs a sophisticated RAG system tailored to the unique demands of the federal civil service labor market. By leveraging the U.S. federal government as a prototype due to its size, diversity, and standardized competency frameworks, a replicable methodology has been developed that offers valuable insights for organizations navigating AI integration. The true power of this approach lies in its adaptability across different competency frameworks, allowing organizations to apply NLP and AI techniques with standardized descriptions and ratings. By training domain-specific models using industry-relevant job descriptions and competencies, organizations can capture nuanced linguistic elements essential for their sectors.

We offer this framework as an initial but rigorous prototype. The CAS scores reported here represent a single model's assessments conditioned on a curated scholarly knowledge base, and they should be interpreted as theory-informed estimates rather than definitive predictions. Future work (already underway) will subject these outputs to systematic reliability testing across multiple LLMs, formal validity and quality evaluations grounded in measurement theory, and entailment analysis to assess inferential consistency (Overton, Robison, and Sheneman 2025). That said, the face validity of the results is encouraging: the model's rankings align with intuitions from domain experts and with independent findings in the economics literature on AI-exposed occupations. We are confident that this methodology establishes a productive foundation for a new generation of workforce planning tools, even as we recognize that its full potential will be realized through iterative refinement and expanded validation.

For public administration scholars, this study contributes a replicable methodology for assessing technology's impact on public work at a level of granularity that generic automation-risk indices cannot provide. By centering the analysis on competencies rather than tasks, the framework shifts attention from the ephemeral question of which activities machines can currently perform to the more durable question of how AI interacts with the human capacities that sustain effective governance. Tasks will continue to be reorganized as generative AI evolves; the knowledge, skills, and abilities through which public servants exercise judgment, ensure accountability, and deliver public value are the stable analytical units on which workforce strategy should be built. For practitioners, the framework offers a path beyond alarmist or utopian narratives toward targeted, evidence-based investments in human capital. The federal workforce is facing transformation, and that transformation demands careful, competency-by-competency analysis that takes human expertise seriously as the foundation of public service.

**Table 1. Taxonomy of Generative AI Impacts on Occupational Competencies**

| Aspect | Complementarity | Augmentation | Substitutivity |
|---|---|---|---|
| Objective | Enhance human capabilities using existing KSAs | Transform human KSAs to integrate AI into tasks | Replicate or replace human tasks and roles |
| Approach | AI works alongside humans, enhancing efficiency | AI forces a change in human capabilities to collaborate effectively | Automates tasks traditionally performed by humans |
| Outcome | Increases productivity without fundamentally changing human roles | Human roles evolve, requiring new skills to work alongside AI | May lead to job displacement in routine tasks |
| Focus | Efficiency and collaboration | Evolution of human cognition and skills to integrate AI | Functional equivalence to human intelligence |
| Application | Tasks benefiting from enhanced efficiency without cognitive change | Tasks requiring both human cognitive evolution and AI capabilities | Effective for routine, repetitive tasks |

## Table 2. Descriptive Statistics for AI Impact Scores: Overall Sample

| Dimension/KSA | Mean | SD | Min | Q1 | Median | Q3 | Max |
|---|---|---|---|---|---|---|---|
| Comp. – Knowledge | 3.78 | 0.56 | 2.00 | 3.00 | 4.00 | 4.00 | 5.00 |
| Comp. – Skills | 3.39 | 0.97 | 1.00 | 3.00 | 3.00 | 4.00 | 5.00 |
| Comp. – Abilities | 3.52 | 0.77 | 1.00 | 3.00 | 4.00 | 4.00 | 5.00 |
| Aug. – Knowledge | 3.42 | 0.63 | 2.00 | 3.00 | 3.00 | 4.00 | 5.00 |
| Aug. – Skills | 3.22 | 0.92 | 1.00 | 2.00 | 3.00 | 4.00 | 5.00 |
| Aug. – Abilities | 3.29 | 0.76 | 1.00 | 3.00 | 3.00 | 4.00 | 5.00 |
| Sub. – Knowledge | 2.45 | 0.56 | 1.00 | 2.00 | 2.00 | 3.00 | 4.00 |
| Sub. – Skills | 2.42 | 0.95 | 1.00 | 2.00 | 2.00 | 3.00 | 5.00 |
| Sub. – Abilities | 2.26 | 0.71 | 1.00 | 2.00 | 2.00 | 3.00 | 4.00 |

## Table 3. Descriptive Statistics by Occupational Family

| Family / Dimension | Mean | SD | Min | Q1 | Median | Q3 | Max |
|---|---|---|---|---|---|---|---|
| White Collar – Comp. | 3.74 | 0.69 | 1.00 | 3.00 | 4.00 | 4.00 | 5.00 |
| White Collar – Aug. | 3.49 | 0.70 | 1.00 | 3.00 | 4.00 | 4.00 | 5.00 |
| White Collar – Sub. | 2.52 | 0.73 | 1.00 | 2.00 | 2.00 | 3.00 | 5.00 |
| TCL – Comp. | 3.11 | 0.86 | 1.00 | 3.00 | 3.00 | 4.00 | 5.00 |
| TCL – Aug. | 2.87 | 0.80 | 1.00 | 2.00 | 3.00 | 3.00 | 5.00 |
| TCL – Sub. | 2.04 | 0.72 | 1.00 | 2.00 | 2.00 | 3.00 | 4.00 |

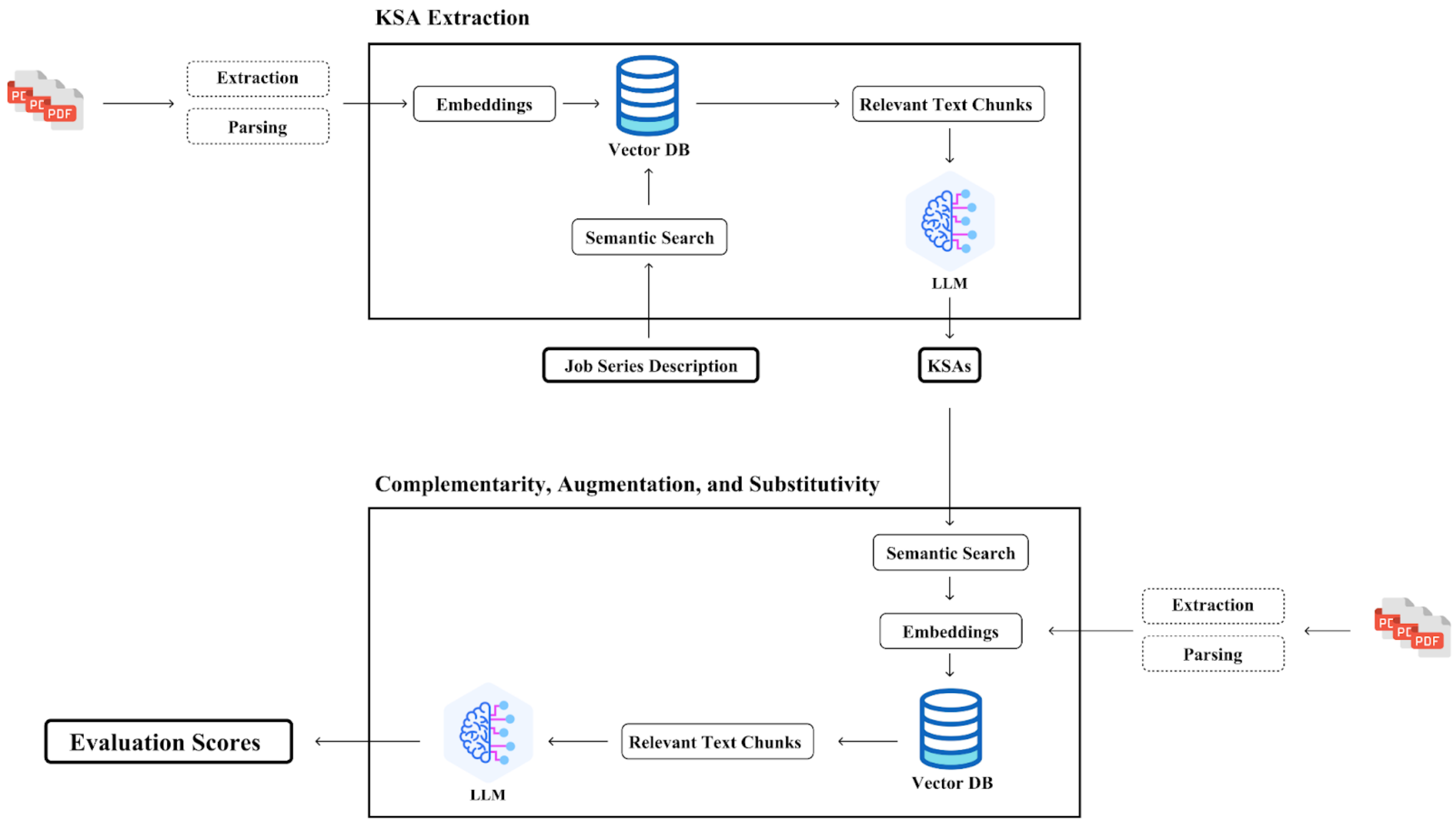

**Figure 1.** Multi-Stage RAG-LLM for Competency Extraction and AI Impact Assessment

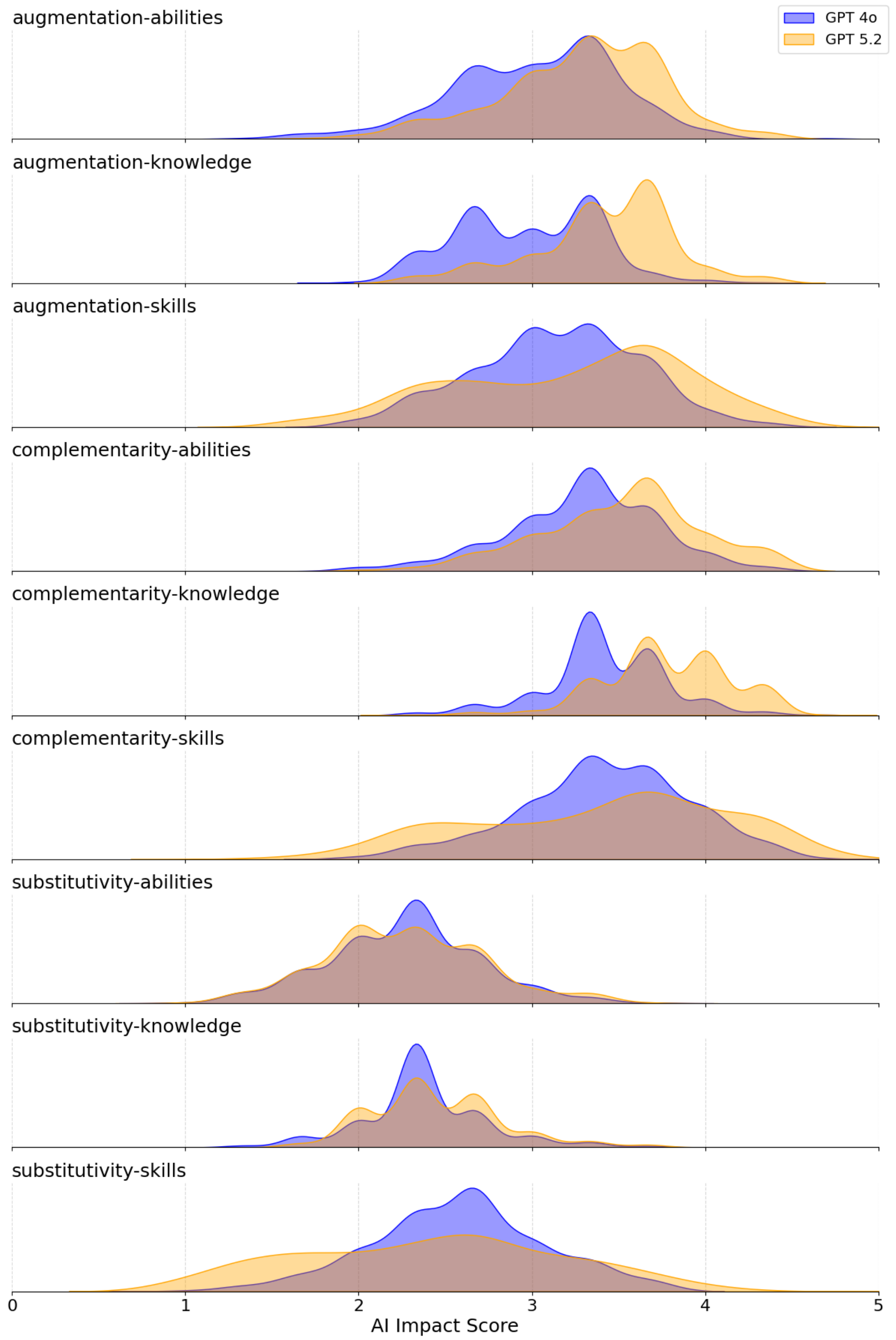

**Figure 2.** Distribution of AI Impact Scores Across KSAs and Occupational Categories

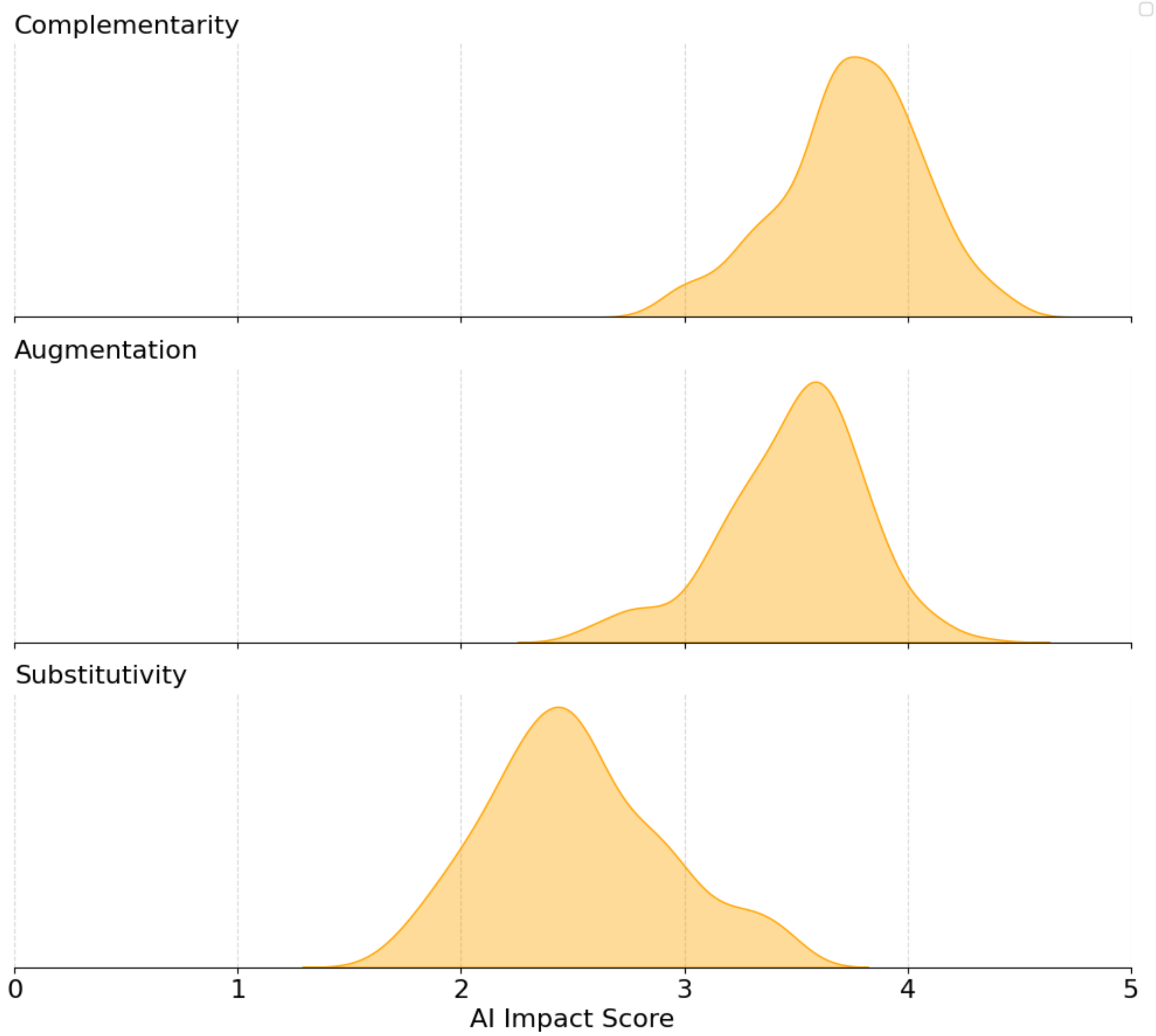

**Figure 3:** Distribution of AI Impact Categories for White Collar Jobs by Competency

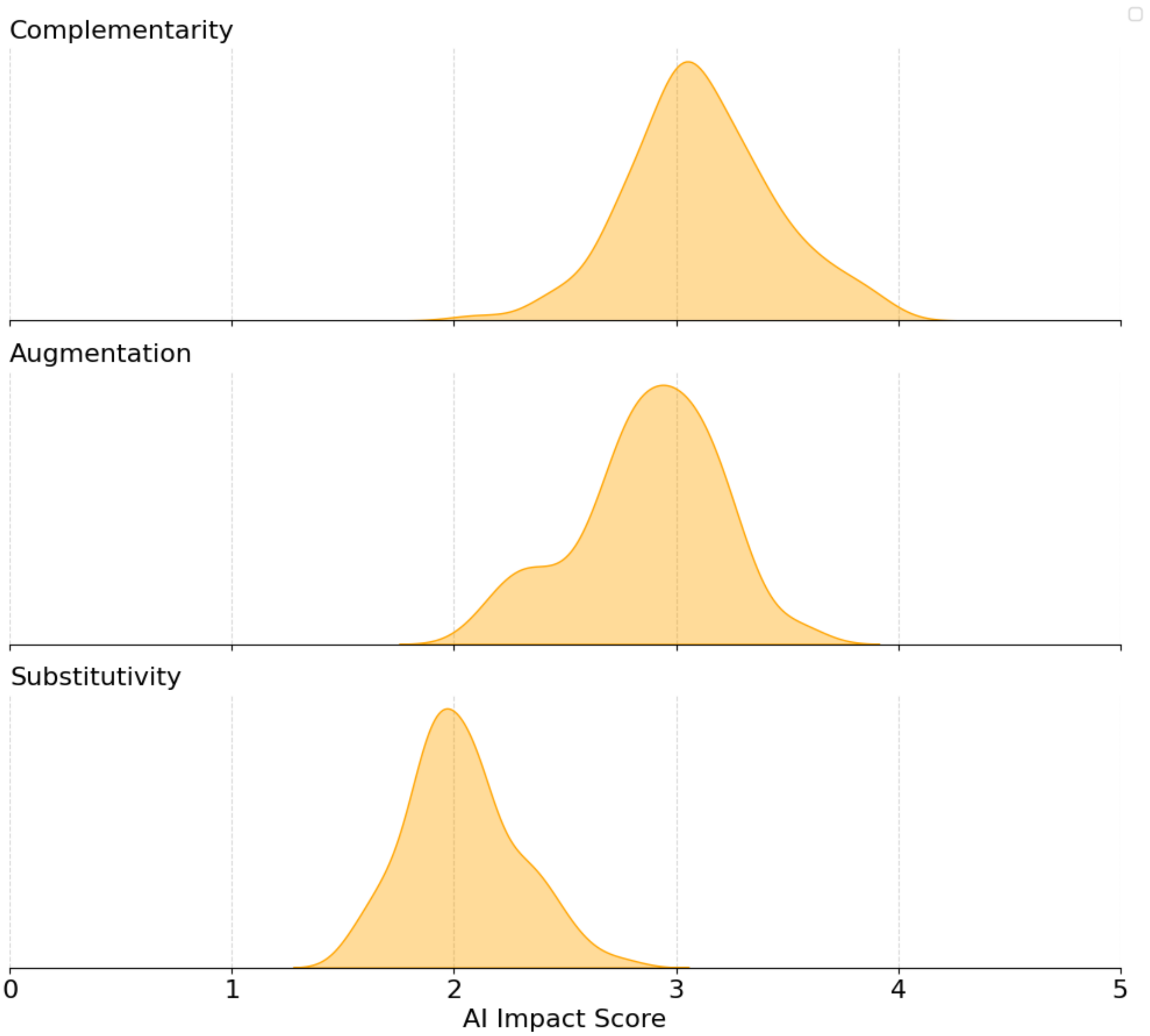

**Figure 4:** Distribution of AI Impact Categories for Trade, Craft, and Labor Jobs by Competency

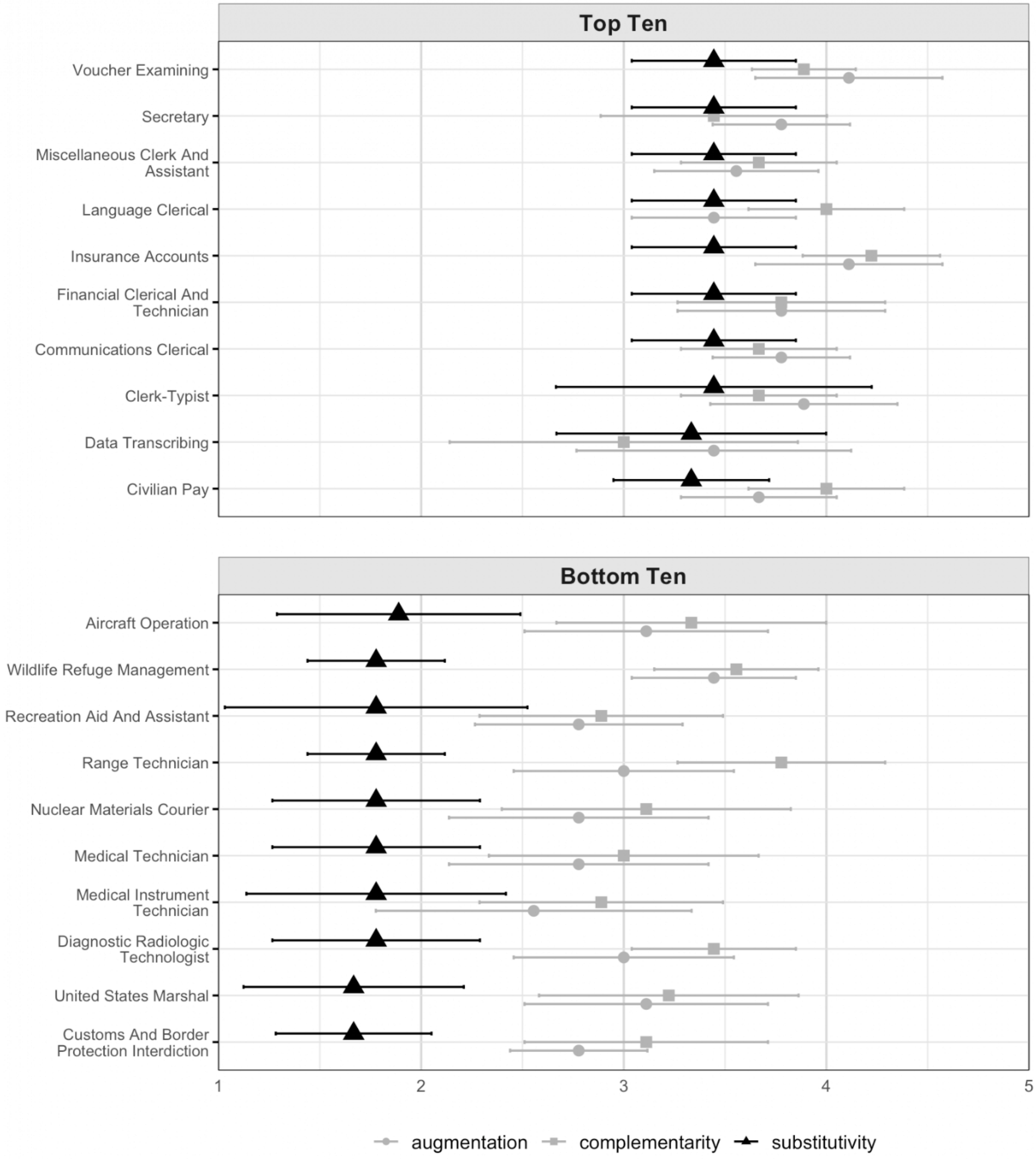

**Figure 5:** Ranking of White-Collar Occupations by Substitutivity Score

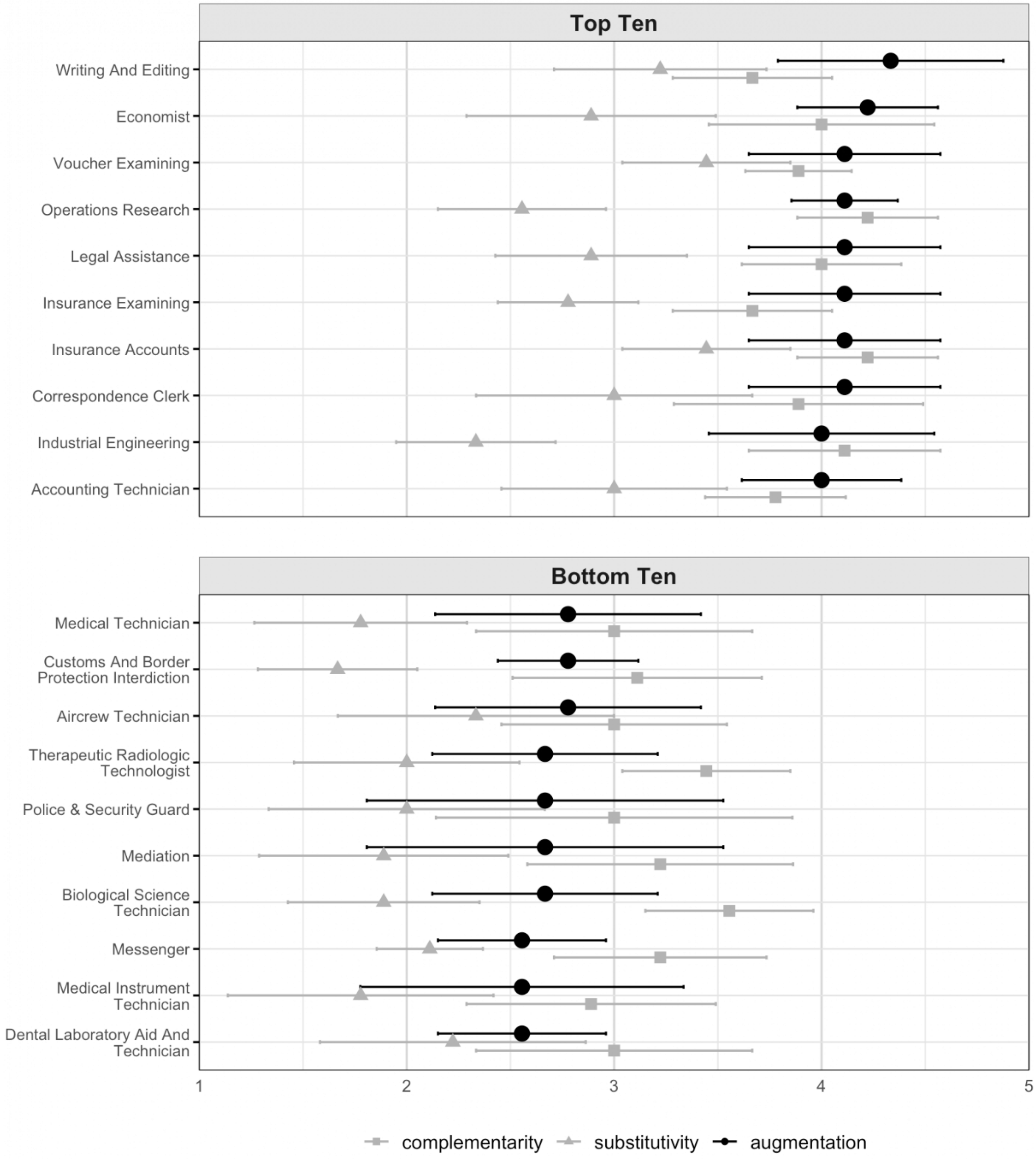

**Figure 6:** Ranking of White-Collar Occupations by Augmentation Score

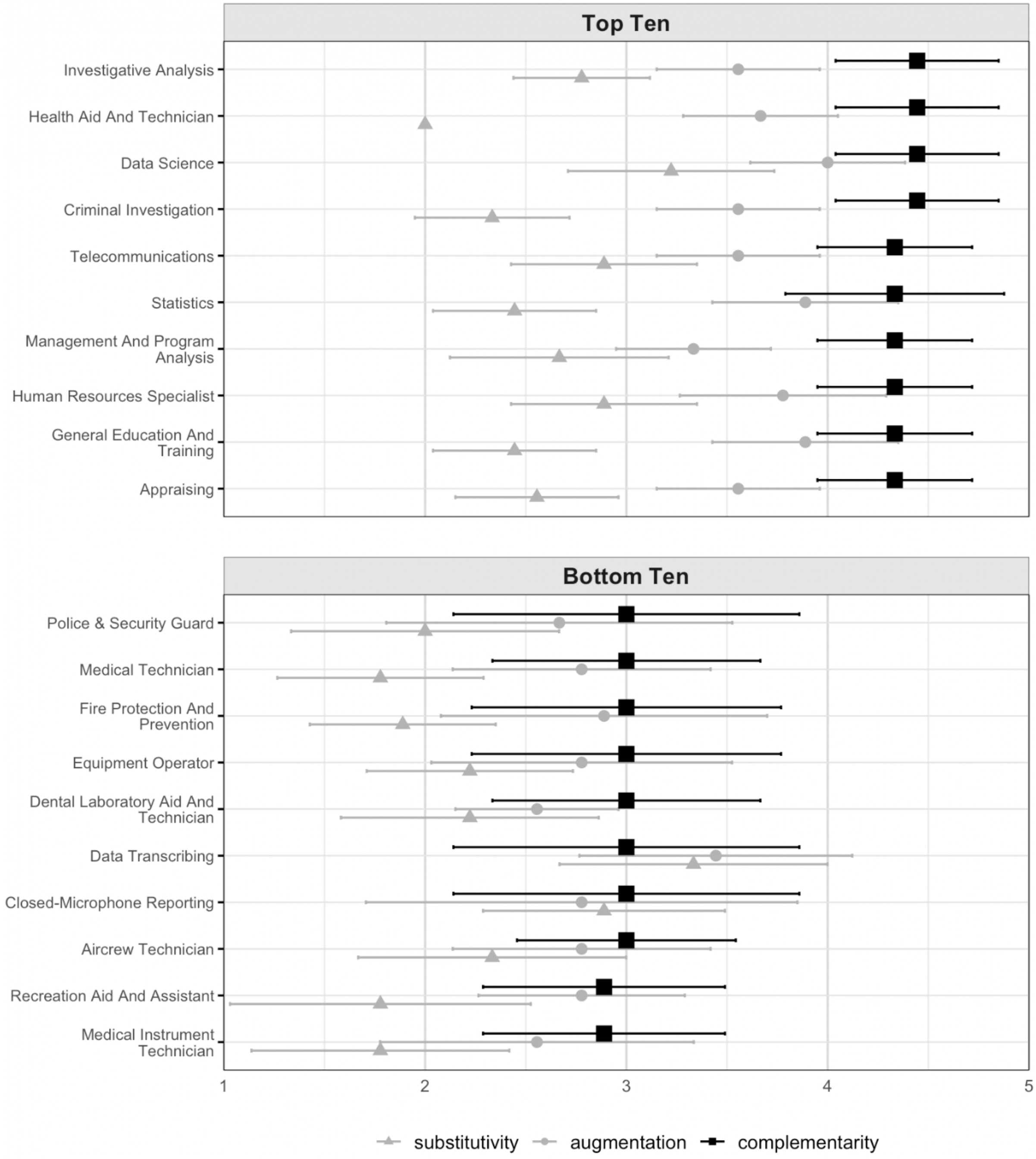

**Figure 7:** Ranking of White-Collar Occupations by Complementarity Score

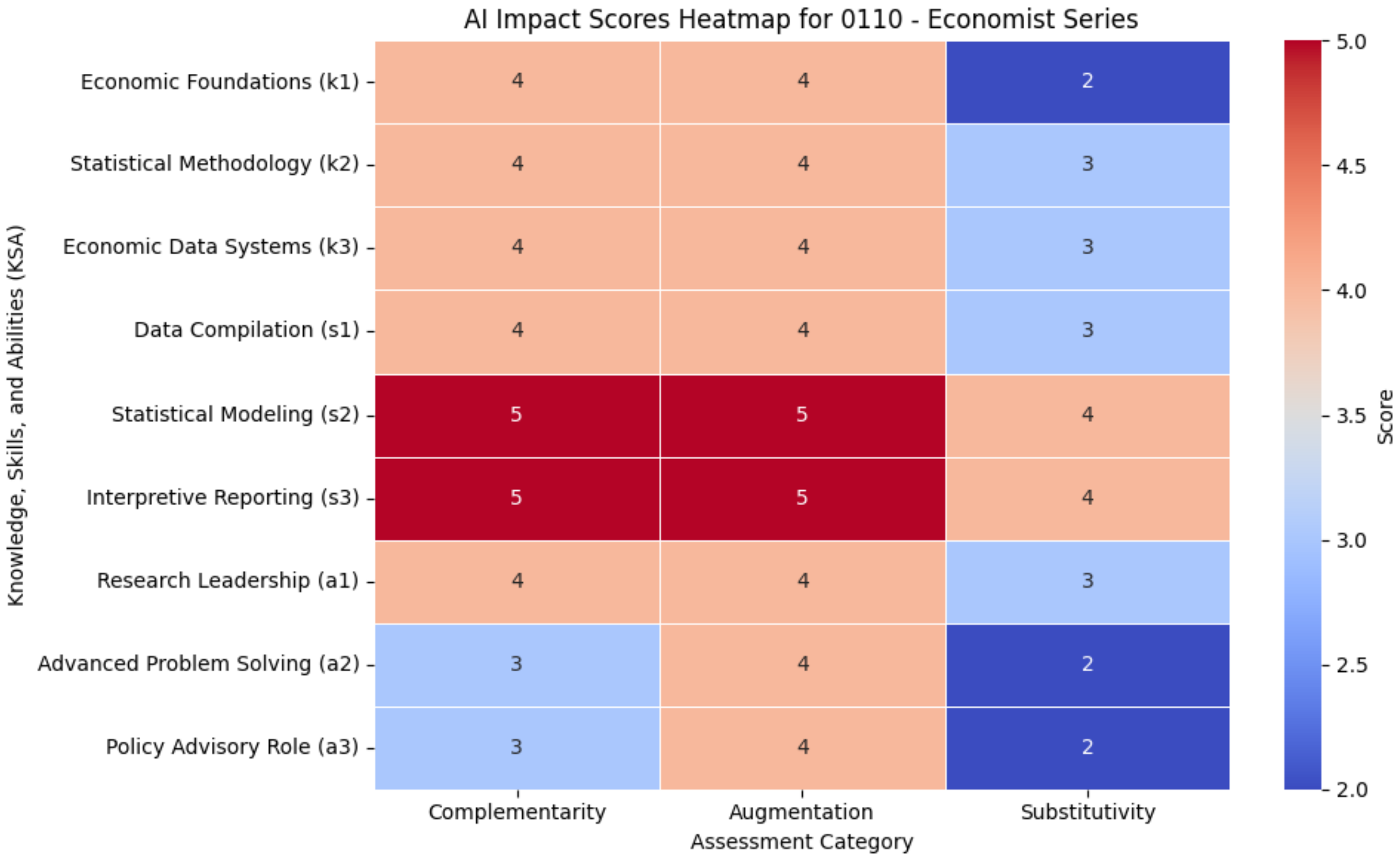

**Figure 8.** Heatmap of CAS Scores Across the Economist Series (0110) KSAs

# Appendix A. Prompts

## A.1 Prompt 1: KSA Extraction

*You are a sophisticated analyst skilled in interpreting job descriptions to identify and categorize the critical knowledge, skills, and abilities (KSAs) essential to job performance. For the given job series, extract the three most critical of each category of knowledge, skill, and ability so that there are nine in total.*

*Knowledge refers to an organized body of information, usually of a factual or procedural nature, which, if applied, makes adequate performance on the job possible. A body of information applied directly to the performance of a function.*

*Skill refers to the proficient manual, verbal or mental manipulation of data or things. Skills can be readily measured by a performance test where quantity and quality of performance are tested, usually within an established time limit. Examples of proficient manipulation of things are skill in typing or skill in operating a vehicle. Examples of proficient manipulation of data are skill in computation using decimals; skill in editing for transposed numbers, etc.*

*Ability refers to the power to perform an observable activity at the present time. This means that abilities have been evidenced through activities or behaviors that are similar to those required on the job, e.g., the ability to plan and organize work. Abilities are different from aptitudes. Aptitudes are only the potential for performing the activity.*

*Stick closely to the original text in the job description, paraphrasing only when needed, and emphasize relevance and precision in each selection.*

### A.2 Prompt 2: AI Impact Assessment (Complementarity)

*You are an expert analyst specializing in evaluating and predicting the impact of generative AI over the next five years on occupations within the United States federal government. You evaluate the impact of generative AI on KSAs for different job series. The overall impact of generative AI can be assessed by different underlying dimensions. For the following task, you will evaluate strictly on the dimension of Complementary Intelligence.*

*Definition of Complementary Intelligence: Complementary intelligence refers to generative AI that works alongside humans by enhancing human capabilities through distinct AI strengths without replacing human labor. The focus is on complementing, rather than replicating, human cognitive knowledge, skills, abilities (KSAs), or tasks.*

*Task: For each given Knowledge, Skill, and Ability, assess how generative AI will impact the job role in terms of complementarity within a five-year window from today.*

*Instructions: Provide a score from 1 to 5 for each KSA using the following scale: 1 (No Complementarity), 2 (Low Complementarity), 3 (Moderate Complementarity), 4 (High Complementarity), 5 (Very High Complementarity).*

*For each score, write a brief explanation (1-2 sentences) justifying your assessment, focusing on how generative AI supports human competencies without replacing them. Do not use markdown formatting, bullet points, or any other special characters in your response. Strictly adhere to the JSON format provided.*

### A.3 Prompt 3: AI Impact Assessment (Augmentation)

*You are an expert analyst specializing in evaluating and predicting the impact of generative AI over the next five years on occupations within the United States federal government. You evaluate the impact of generative AI on KSAs for different job series. The overall impact of generative AI can be assessed by different underlying dimensions. For the following task, you will evaluate strictly on the dimension of Augmented Intelligence.*

*Definition of Augmented Intelligence: Augmented intelligence refers to the impact of AI that requires a transformation of human capacities. It involves integrating AI in a way that necessitates changes in knowledge, skills, abilities (KSAs), or tasks to improve decision-making while still maintaining human involvement. The focus is on the extent of necessary evolution of human cognitive KSAs to integrate AI.*

*Task: For each of the given Knowledge, Skills, Abilities, assess how generative AI will impact the job role in terms of augmented intelligence within a five-year window from today.*

*Instructions: Provide a score from 1 to 5 for each KSA: 1 (No Augmentation) to 5 (High Augmentation). For each score, write a brief explanation (1-2 sentences) justifying your assessment, focusing on how generative AI requires changes in human input. Do not use markdown formatting, bullet points, or any special characters in your response. Strictly adhere to the JSON format provided.*

### A.4 Prompt 4: AI Impact Assessment (Substitutivity)

*You are an expert analyst specializing in evaluating and predicting the impact of generative AI over the next five years on occupations within the United States federal government. You evaluate the impact of generative AI on KSAs for different job series. The overall impact of generative AI can be assessed by different underlying dimensions. For the following task, you will evaluate strictly on the dimension of Substitutive Intelligence.*

*Definition of Substitutive AI: Substitutive AI refers to artificial intelligence systems designed to fully replicate or replace functions traditionally performed by humans. In this approach, human involvement is minimized or eliminated as AI takes over functional responsibilities. Substitutive AI aims for automation that mirrors human capabilities without requiring human input, effectively achieving functional equivalence to human intelligence in specific functions.*

*Task: For each of the given Knowledge, Skills, Abilities, assess how generative AI will impact the job role in terms of substitutivity within a five-year window from today.*

*Instructions: Provide a score from 1 to 5 for each KSA: 1 (No Substitution) to 5 (Full Substitution). For each score, write a brief explanation (1-2 sentences) justifying your assessment, focusing on the likelihood of AI fully automating or replacing the human input. Do not use markdown formatting, bullet points, or any special characters in your response. Strictly adhere to the JSON format provided.*

## Appendix B. Trade, Craft, and Labor Occupational Rankings

TCL occupations make up approximately 10 percent of federal government positions. Figures B.1 through B.3 rank TCL occupational series by average scores for each dimension.

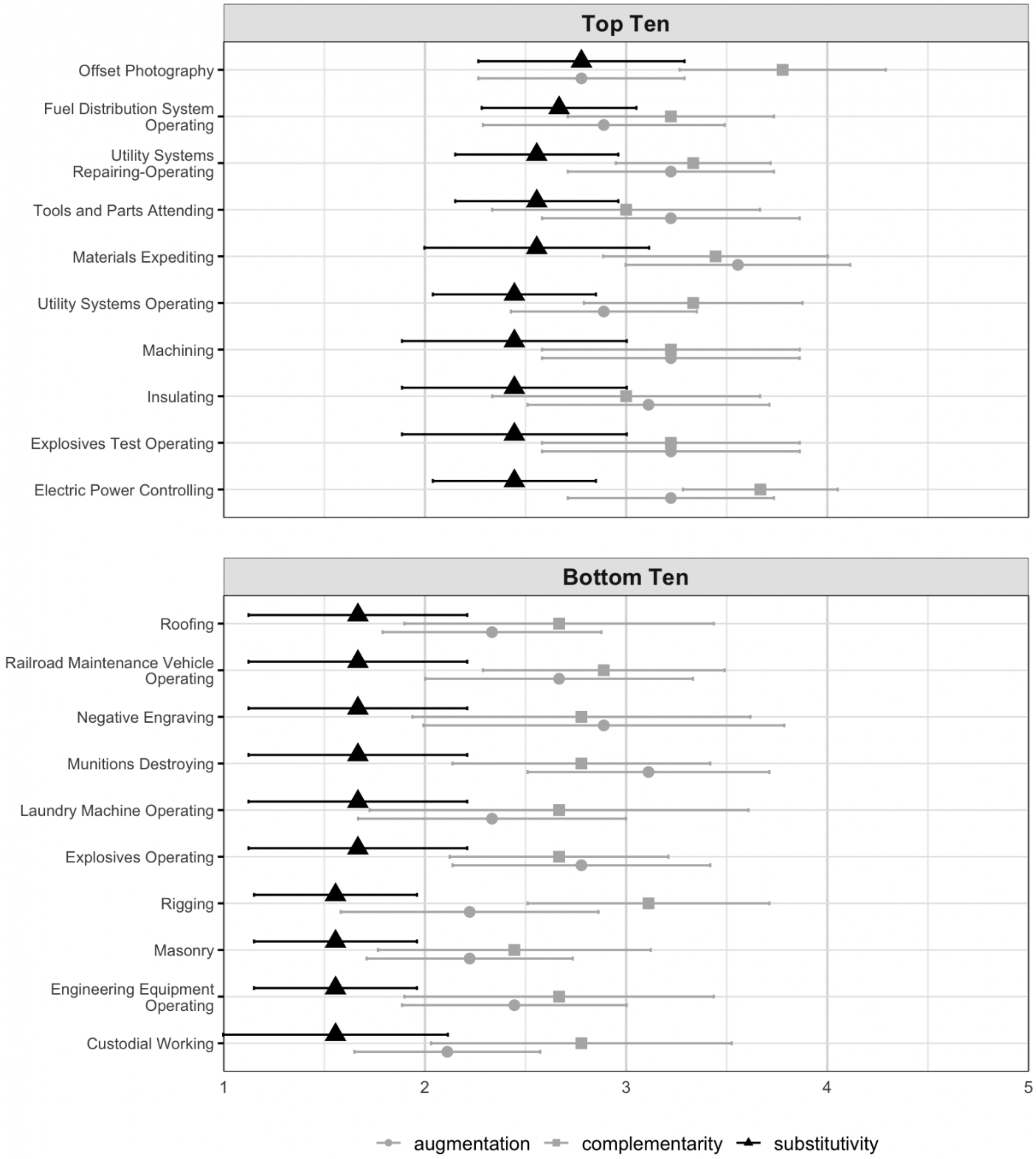

Among TCL occupations, offset photography, fuel distribution system operating, and utility systems repairing-operating rank highest in substitutivity, reflecting procedural, repetitive tasks amenable to automation. Masonry, engineering equipment operating, and custodial working exhibit the lowest substitutivity, as these roles demand physical dexterity, situational judgment, and hands-on expertise.

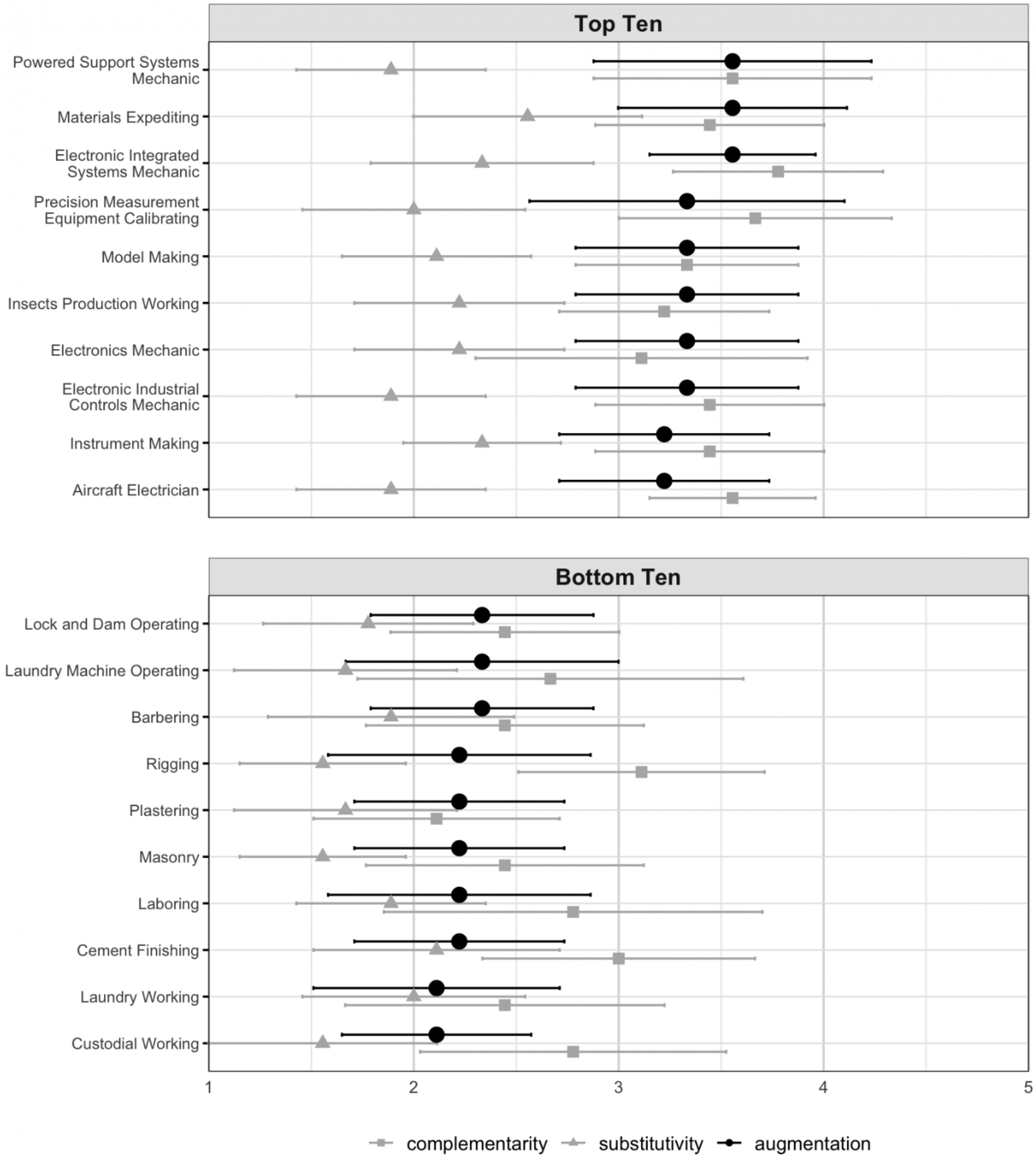

For augmentation, powered support systems mechanic, materials expediting, and electronic integrated systems mechanic rank highest, reflecting roles where AI-generated diagnostics can transform how workers approach technical tasks. Cement finishing, laundry, and custodial roles rank lowest.

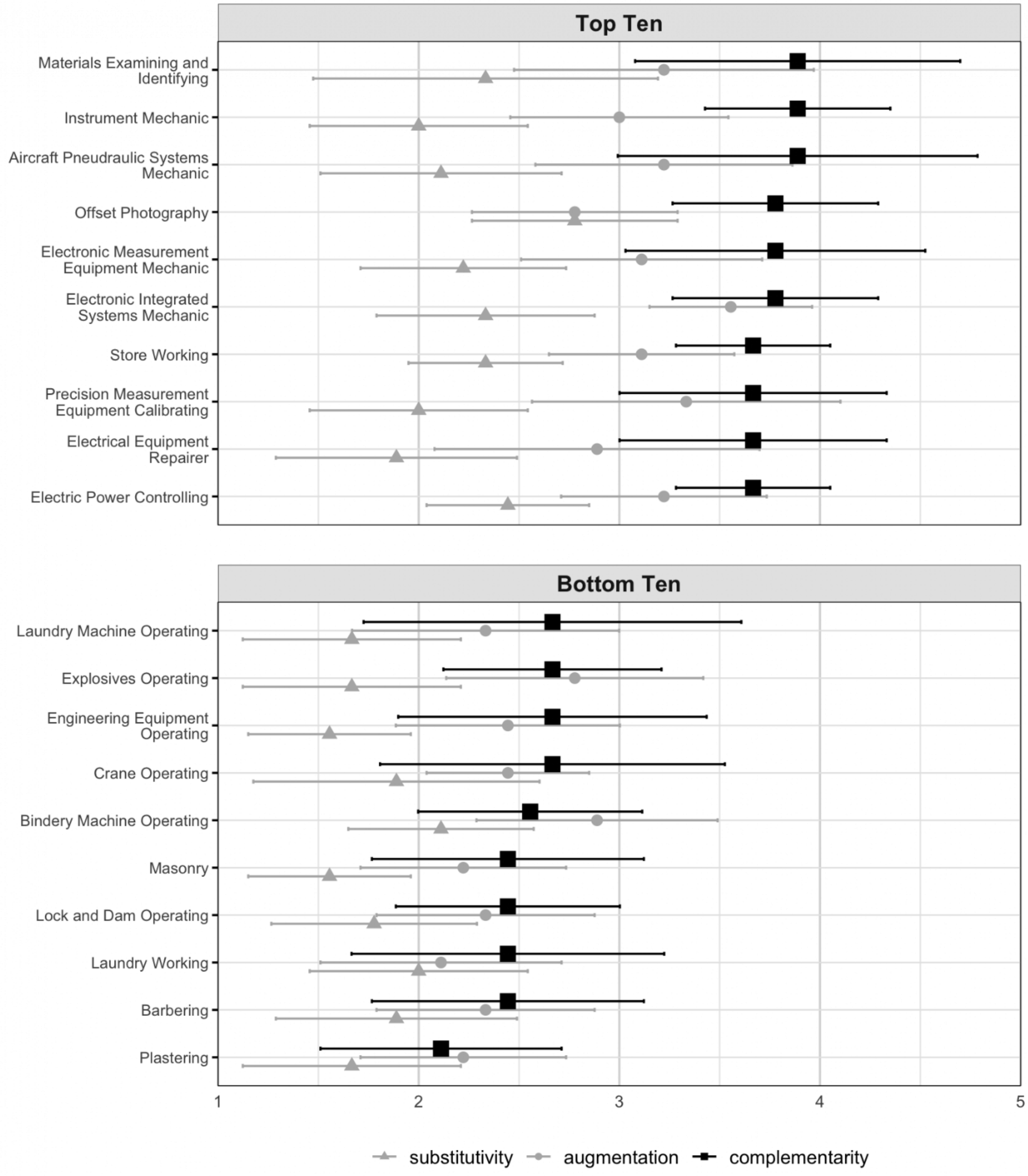

Materials examining and identifying, instrument mechanic, and aircraft pneudraulic systems mechanic exhibit the highest complementarity among TCL roles. Laundry, barbering, and plastering show the lowest.

## Appendix C. Validation: AI's Impact on Economists

To assess consistency with the emerging scholarship, the Economist series results are compared with recent insights on generative AI's impact on economics.

Recent studies underscore the transformative role of AI in economic data-intensive tasks. Korinek (2023) highlights that generative AI tools can automate numerous micro-tasks in research, such as data analysis and coding assistance, thereby allowing economists to focus more on interpreting results and applying insights. Desai (2023) emphasizes that machine learning models excel at processing large volumes of unstructured data, capturing complex patterns, and enhancing predictive accuracy, which complements traditional econometric approaches. These capabilities align with the high complementarity and augmentation scores observed for statistical theory and data analysis, as AI streamlines computations and data handling, enabling economists to delve deeper into strategic analysis.

The integration of AI into economic modeling and policy-making presents nuanced challenges. Marwala and Hurwitz (2017) discuss that while AI techniques can enhance economic models by providing more accurate simulations and trend analyses, human judgment remains paramount for refining these models and contextualizing outcomes within theoretical frameworks. This perspective supports the moderate substitutivity scores for economic modeling, indicating that AI serves as a valuable tool but cannot fully replace the strategic insights provided by economists. The Congressional Budget Office (2024) notes that although AI has the potential to transform service delivery in both business and government sectors, its ability to substitute human expertise is limited in areas requiring deep contextual understanding and adaptability, such as interpreting economic principles and understanding institutional frameworks. This limitation is reflected in the lower substitutivity scores for these knowledge areas.

Occupations that are data-intensive, such as those in data science or physics, are among the most impacted. AI has proven particularly effective at enhancing roles that require large-scale data processing, predictive analytics, and computational problem-solving. Agrawal, Gans, and Goldfarb (2018) underscore that AI excels in tasks involving pattern recognition and data analysis, thereby automating routine aspects of information management. In these fields, AI augments rather than replaces human labor, allowing workers to focus on interpreting AI-generated results and applying insights. These comparisons validate both the construct validity of the three-dimensional framework and the discriminant validity of individual KSA-level assessments, demonstrating that the model's distinctions between complementarity, augmentation, and substitutivity correspond meaningfully to observed patterns in how AI is reshaping professional practice.